\documentclass[prx,11pt,nofootinbib,superscriptaddress,longbibliography]{revtex4-2}
\usepackage{graphicx}
\usepackage{amsmath}
\usepackage{amstext}
\usepackage{amssymb}
\usepackage{xfrac}
\usepackage[colorlinks,citecolor=blue]{hyperref}
\usepackage{graphicx}
\usepackage{amsmath}
\usepackage{amstext}    
\usepackage{amssymb}
\usepackage{amsfonts}
\usepackage{longtable,booktabs}
\usepackage{hyperref}
\usepackage{url}
\usepackage{subfigure}
\usepackage{dsfont}
\usepackage{amsbsy}
\usepackage{dcolumn}
\usepackage{amsthm}
\usepackage{bm}
\usepackage{esint}
\usepackage{multirow}
\usepackage{hyperref}
\usepackage{cleveref}
\usepackage{rotating}
\usepackage{mathrsfs}
\usepackage{amsfonts}
\usepackage{amsbsy}
\usepackage{dcolumn}
\usepackage{bm}
\usepackage{multirow}
\usepackage{color}
\usepackage{extarrows}
\usepackage{datetime}
\usepackage{comment}
\usepackage[super]{nth}
\usepackage{booktabs}
\usepackage{natbib}[numbers]

\newtheorem{statement}{Statement}

\pdfoutput=1
\usepackage[T1]{fontenc}
\usepackage[latin9]{inputenc}
\usepackage{times}

 \providecommand{\tabularnewline}{\\}

\makeatother

\begin{document}
  
\title{{\Large  Many-body physics of  spontaneously broken higher-rank symmetry: from fractonic superfluids to dipolar Hubbard model}}
 
\author{Shuai A. Chen}
\email{chsh@ust.hk}\thanks{\scriptsize Present address: Department of Physics, Hong Kong University of Science and Technology, Clear Water Bay, Hong Kong, China.}

\affiliation{Guangdong Provincial Key Laboratory of Magnetoelectric
Physics and Devices, Sun Yat-sen University, Guangzhou, 510275, China }
\affiliation{Department of Physics, Hong Kong University of Science and Technology,
Clear Water Bay, Hong Kong, China}
\author{Peng Ye}
\email{yepeng5@mail.sysu.edu.cn}
\affiliation{Guangdong Provincial Key Laboratory of Magnetoelectric
Physics and Devices, Sun Yat-sen University, Guangzhou, 510275, China }
\affiliation{School of Physics, Sun Yat-sen University, Guangzhou, 510275, China }

\begin{abstract}

Fractonic superfluids are exotic phases of matter in which bosons are subject to mobility constraints, resulting in features beyond those of conventional superfluids. These exotic phases arise from the spontaneous breaking of higher-rank symmetry (HRS) in many-body systems with higher-moment conservation, such as dipoles, quadrupoles, and angular moments. The aim of this paper is to introduce  exciting  developments on the theory of spontaneous symmetry breaking in such systems, which we refer to as ``many-fracton systems''. More specifically,  we introduce exciting progress on general aspects of HRS, minimal model construction, realization of 
symmetry-breaking ground states, order parameter, off-diagonal long-range order (ODLRO),  Noether currents with continuity equations, Gross-Pitaevskii equations, quantum fluctuations, Goldstone modes, specific heat, generalized Mermin-Wagner theorem, critical current, Landau criterion, symmetry defects, and Kosterlitz-Thouless (KT)-like physics, hydrodynamics, and dipolar Hubbard model realization. This paper is concluded with several future directions.\footnote{{\scriptsize Acknowledgements: This  research  was conducted during S.A.C.'s visit to Guangdong Provincial Key Laboratory of Magnetoelectric Physics and Devices (LaMPad) in Sun Yat-sen University, and  was  fully supported by  the Open Project of LaMPad under Grant No. 2022B1212010008, NSFC Grant No.~12074438 \& 11847608, and Guangdong Basic and Applied Basic Research Foundation under Grant No.~2020B1515120100. The authors acknowledge Jian-Keng Yuan, Hongchao Li, Meng-Yuan Li, Chengkang Zhou, Zheng Yan, Zi Yang Meng,  T.K. Ng, K.T. Law, Y.B. Yang, Meng Cheng, Juven Wang, Ruizhi Liu, Yixin Xu, et.al. for   valuable discussions in the past years. P.Y. would like to thank Gang Chen for the   hospitality and the arrangement of JITCP seminar at the University of Hong Kong, where the main part of this draft was also  completed.}\bigskip
}

\end{abstract}

\maketitle
\tableofcontents{}

\section{Introduction}


Symmetry is a fundamental concept in theoretical physics that plays a central role in describing the behavior of physical systems. The idea of \textit{spontaneous symmetry breaking} (SSB) has been observed in various contexts, ranging from particle physics to condensed matter physics. It is closely related to the description of phase transitions and the Standard Model of our universe, which unifies classical criticality within the framework of Landau's paradigm and fundamental interactions within a large non-Abelian group \citep{volovik2003}. SSB is mathematically described by group theory, and it continues to be a focus of research with far-reaching impacts.
Although tremendous progress has been made in exploring quantum matters beyond SSB, such as topologically ordered phases exemplified by the fractional quantum Hall effect and free-fermion topological phases exemplified by quantum spin Hall insulators, the idea of SSB remains relevant and continues to inspire new research directions. For example, SSB in strongly-correlated systems (e.g., cuprates) \citep{doi:10.1146/annurev-conmatphys-031119-050711} exhibits a diverse array of phenomena that are still not fully understood. Additionally, efforts continue to be made to generalize the theoretical framework of SSB itself, opening up new avenues of inquiry for researchers in theoretical physics.

 In this paper, we aim to introduce  recent progress on the generalization of SSB by means of `new' global symmetry called ``higher-rank symmetry'' (HRS)\footnote{While there are    terminologies of many variants,  here by higher-rank, we mean that, upon gauging this symmetry, the resulting gauge field is  a tensor of higher-rank and not a differential form in the continuum limit.}. This line of thinking 
 was recently initiated from the   field of    fracton topological order or, more loosely speaking, fracton
physics~(see, e.g.,  Refs.~\cite{Nandkishore2019,2020Fracton} and   references therein).   Fracton topological orders
 represent a class of gapped phases of matter that are  characterized by noise-immune ground state degeneracy
that depends on the system size in a non-trivial way on a compact
manifold. 
Topological excitations of fracton topological order include strange particles such as fractons,
subdimensional particles \citep{Chamon05,Haah2011,Vijay2015,Vijay2016,Pretko2017} and also exotic spatially extended excitations studied in \citet{ye19a,limengyuan2,2023PhRvB.107k5169L}.
A remarkable feature of these excitations is their restricted mobility due to the absence of arbitrarily deformable   string operators in the stabilizer code models. In other words, the locations of fractons cannot be locally changed by any local operators, which is expected to  the potential realization of robust  quantum memory. In addition to fractons, in a class of type-I fracton lattice models, there are point-like excitations that can move within subdimensional manifolds, called ``subdimensional particles''. For example,
in the X-cube lattice model \citep{Chamon05,Haah2011,Vijay2015,Vijay2016,PhysRevResearch.4.033111},
\textit{lineons} can move along certain one-dimensional directions and \textit{planeons}
can move within two-dimensional planes. In addition to the X-cube model, there are also many other variants of fracton topological order, some of which are dual to symmetry-protected topological phases with subsystem symmetry \citep{PhysRevB.98.035112,2018PhRvB..98w5121D,2022PhRvB.106h5113B,2021PhRvB.103c5148S,PhysRevB.99.205109,PhysRevB.105.245122,2019Quant3142S,2020AnPhy.41668140Y,2018PhRvA..98b2332D,Shirley2019,PhysRevResearch.2.012059,PhysRevB.106.214428}.

Instead of ``topological excitations'', one can also regard the strange
particles, i.e., fractons, lineons, and planeons, as microscopically constituent particles\footnote{
Unless otherwise specified, our primary focus in this paper is on bosonic systems. However, we acknowledge that fermions also play a crucial role in many-body systems and their study is equally important.}
of some underlying many-body systems (dubbed ``many-fracton systems''), 
which leads to potentially unconventional many-body physics. Within this context, the number of fractons is  no longer dilute  but reaches the thermodynamical limit. 
Remarkably, in such many-body systems, a natural way to implementation of mobility restriction is to enforce higher-moment (e.g., dipoles, angular moments, and quadrupoles) conservation on top of the total charge (i.e., particle number) conservation \citep{Pretko2018,2018arXiv180601855B,PhysRevX.9.031035,Seiberg2019arXiv1909}, which, by Noether's theorem, is equivalent to the requirement of exotic continuous symmetry (i.e., HRS in the present paper) that keeps the underlying Hamiltonian (or Lagrangian) invariant.  
Following the well-accepted spirit of symmetry implementation in quantum theory, the conserved quantities, e.g., total charge, dipoles, and angular moments, play the role of the HRS generators. Therefore, unitary operators of symmetry transformations transform matter field operators in a coordinate dependent fashion, which is in sharp distinction from  from the conventional global symmetry, e.g., $U(1)$ symmetry of particle number conservation. Superficially, it looks like a gauge transformation but there are two explicit differences. First, the Hamiltonian is invariant once  HRS acts on matter fields (e.g., fractons, lineons, etc.) without the need of extra gauge degrees of freedom. Second, the spatial-dependence is rather restricted, such as linear dependence, while  gauge transformations  are in principle allowed to be arbitrarily dependent on coordinates. 
By noting that there have been great triumphs in the conventional
spontaneously symmetry-breaking phase, we are motivated to ask if
there is novel many-body physics by instead considering HRS. For example,
in the framework of HRS, what can we expect for the superfluid-like
phase, off-diagonal long-range order (ODLRO), generalized Mermin-Wagner theorem, Gross-Pitaevskii equations, hydrodynamical theory, critical current, Landau criterion, lattice realization, 
symmetry defects, and Kosterlitz-Thouless (KT)-like physics?

Along this line of thinking,
   a concrete model study on an SSB phase with higher moment conservation, dubbed 
\textit{fractonic superfluid phase}, was performed in  \citet{2020PhRvR2b3267Y} [see also arXiv:1911.02876 (2019)] and subsequently in \citet{2020arXiv201003261C,2021PhRvR3d3176L,2022arXiv220108597Y,2022ChPhL..39e7101Y}. 
Many exotic
phenomena beyond the conventional symmetry-breaking phases have been
identified in various channels.  Table~\ref{Tab}  provides a comparison between a conventional superfluid phase (denoted as $d\mathsf{SF}^{d}$,
see~\citet{2020arXiv201003261C}), fractonic superfluid phase
(denoted as $d\mathsf{SF}^{0}$; see~\citet{2020PhRvR2b3267Y})
via condensing fractons, and a fractonic superfluid phase (denoted
as $d\mathsf{SF}^{1}$) via condensing lineons.  
For notation convenience, we apply the notation 
$d\mathsf{SF}^{n}$ (``$\mathsf{SF}$'' stands for ``superfluid'') 
in~\citet{2020arXiv201003261C}
to represent a superfluid phase in $d$ spatial dimensions via condensing
subdimensional particles of dimension-$n$ ($0\leq n\leq d$). 
For instance, a fractonic superfluid phase arises from condensing fractons,
i.g. fully localized particles, which can be denoted by $d\mathsf{SF}^{0}$
while the $d\mathsf{SF}^{d}$ describes a conventional superfluid
phase where bosons are free to move in the whole space.

At present, it remains topics of active 
exploration on quantum systems with higher moment conservation and  higher-rank symmetry studied by the community from many different strategies ranging from condensed matter to high energy, see, e.g.,  \citet{2019arXiv191101804W,2021CmPhy...4...44D,2022PhRvR...4b3151G,2022PhRvB.106f4511L,2022arXiv221002470L,2022PhRvB.105o5107S,2022arXiv220501132J,PhysRevB.106.245125,2023arXiv230110782B,2022arXiv221011072Z,2023arXiv230409596A,2023arXiv230409852J,PhysRevResearch.4.033186,2023arXiv230403276H,2023arXiv230412911A} and other references appearing in the maintext of this paper.   
\begin{table*}
\centering
\caption{Comparison between a conventional superfluid phase (denoted as $d\mathsf{SF}^{d}$,
see~\citet{2020arXiv201003261C}), fractonic superfluid phase
(denoted as $d\mathsf{SF}^{0}$; see~\citet{2020PhRvR2b3267Y})
via condensing fractons, and a fractonic superfluid phase (denoted
as $d\mathsf{SF}^{1}$) via condensing lineons. In these three types
of superfluids, the condensed particles are, respectively, usual bosons
of full mobility, fractons without any mobility, and lineons with
partial mobility. Vortex excitations in $2$D form a hierarchy where
$\ell,\ell_{1},\ell_{2}$ denote winding numbers and $p,p_{1},p_{2}$
are quantized as momenta with $\varphi(\mathbf{x})$ being the relative
angle of site $\mathbf{x}$ to the vortex core. 
From \citet{2020arXiv201003261C}. }
\label{Tab}%
\begin{tabular}{cccc}
\specialrule{0.1em}{1pt}{0pt}
 & $d\mathsf{SF}^{d}$  & $d\mathsf{SF}^{0}$  & $d\mathsf{SF}^{1}$ \tabularnewline
\hline 
Conserved quantities  & Charge  & Charge, dipole moment  & Charges, angular charge moments \tabularnewline
Order Parameter  & $\sqrt{\rho_{0}}e^{i\theta_{0}}$  & $\sqrt{\rho_{0}}e^{i\left(\theta_{0}+\sum_{a}\beta_{a}x^{a}\right)}$  & $\sqrt{\rho_{0}}e^{i\left(\theta_{a}+\sum_{b}\beta_{ab}x^{a}\right)}\left(\beta_{ab}=-\beta_{ba}\right)$ \tabularnewline
Plane-wave dispersion  & Dispersive  & Dispersionless  & Partially dispersive \tabularnewline
Ground State  & $e^{\int\mathrm{d}^{d}x\sqrt{\rho_{0}}e^{i\theta_{0}}\hat{\Phi}^{\dag}(\mathbf{x})}|0\rangle$  & $e^{\int\mathrm{d}^{d}x\sqrt{\rho_{0}}e^{i\left(\theta_{0}+\sum_{a}\beta_{a}x^{a}\right)}\hat{\Phi}^{\dag}\left(\mathbf{x}\right)}|0\rangle$  & $\prod_{a}e^{\int\mathrm{d}^{d}x\sqrt{\rho_{0}}e^{i\left(\theta_{a}+\sum_{b}\beta_{ab}x^{a}\right)}\hat{\Phi}_{a}^{\dag}(\mathbf{x})}|0\rangle$ \tabularnewline 
Specific capacity heat  & $c_{\upsilon}\varpropto T^{d}$  & $c_{\upsilon}\varpropto T^{\frac{d}{2}}$  & $c_{\upsilon}\varpropto T^{d}$ \tabularnewline
\# of Goldstone modes  & $1$  & $1$  & $d$ \vspace{0.25cm}\tabularnewline
\begin{minipage}{3cm}
Dispersion of Goldstone modes 
\end{minipage} & $\omega\varpropto\left\vert \mathbf{k}\right\vert $  & $\omega\varpropto\left\vert \mathbf{k}\right\vert ^{2}$  & $\omega\varpropto\left\vert \mathbf{k}\right\vert $  \vspace{0.25cm}\tabularnewline
Stable dimension at $T=0$  & $d>1$  & $d>2$  & $d>1$ \vspace{0.25cm} \tabularnewline
Vortex structure in $d=2$  & 
$\ell\varphi\left(\mathbf{x}\right)$  & $\begin{array}{c}\ell\varphi\left(\mathbf{x}\right)\\
 p_{1}x^{1}\varphi(\mathbf{x})+p_{1}x^{2}\log|\mathbf{x}|\\
 p_{2}x^{2}\varphi(\mathbf{x})-p_{2}x^{1}\log|\mathbf{x}|. 
\end{array} $ & $\begin{array}{c}
\theta_{1}:\ell_1\varphi(\mathbf x),\theta_{2}:\ell_2\varphi(\mathbf x)
\\  \theta_{1}:-px^{1}\log|\mathbf{x}|+px^{2}\varphi\left(\mathbf{x}\right) \\
\theta_{2}:-px^{2}\log|\mathbf{x}|-px^{1}\varphi\left(\mathbf{x}\right) 
\end{array}$ \tabularnewline
\specialrule{0.1em}{1pt}{0pt}
\end{tabular}
\end{table*}
In this paper, we will focus on SSB aspects in fractonic superfluids by  briefly introducing  recent  progress on fractonic superfluids and most relevant topics that are shown interests on by the community.   
In Sec.~\ref{section_HRS}, we introduce some facts about HRS for both single-component bosons (Sec.~\ref{section_HRS_single}) and multi-component bosons (Sec.~\ref{section_HRS_multiple}). 
In Sec.~\ref{section_minimal_model}, we will introduce  minimal models that respect HRS and spontaneously  break HRS in ground states.  Two typical examples are introduced, where dipoles (Sec.~\ref{section_minimal_model_dipole}) and angular moments (Sec.~\ref{section_minimal_model_angular}) are conserved respectively. General construction of minimal models are given in Sec.~\ref{section_minimal_model_general}. In Sec.~\ref{section_minimal_model_thm}, recent progress on generalization of Mermin-Wagner theorem is introduced. 
In Sec.~\ref{section_kt}, we focus on symmetry defect of HRS and study the finite temperature phase diagrams of fractonic superfluids via proliferating HRS symmetry defects.  
In Sec.~\ref{section_kt_construction}, we introduce the general theory of construction of HRS symmetry defects. 
In Sec.~\ref{section_kt_rg}, we introduce renormalization group analysis of the hierarchy of Kosterlitz-Thouless transitions. 
In Sec.~\ref{section_hydro}, we introduce the hydrodynamical theory of fractonic superfluids, especially focusing on the appearance of Navier-Stoke-like equations. In Sec.~\ref{section_latt}, we introduce recent progress on lattice realization of many-fracton systems with SSB ground states.
We conclude this paper with an outlook in Sec.~\ref{section_discussion}. 

\section{Higher-rank symmetry and higher-moment conservation}\label{section_HRS}

\subsection{Single-component bosons}\label{section_HRS_single}
Let us begin with a conventional global symmetry, such as a $U(1)$ symmetry
that corresponds to charge conservation. In the second-quantization scheme of many-body systems,
the following canonical bosonic/fermionic commutation relation 
\begin{equation}
[\hat{\Phi}(\mathbf{x}),\hat{\Phi}^{\dagger}(\mathbf{y})]_{\pm}=\delta(\mathbf{x}-\mathbf{y})~,
\label{commutation}
\end{equation}
enables an interpretation to regard the field $\hat\Phi(\mathbf{x})$
to annihilate fermion/boson at the site $\mathbf{x}=(x^1,\cdots,x^d)$ in $d$ spatial dimensions. Then, a global
$U(1)$ symmetry unitary operator can be generated by the charge $\hat{Q}=\int d^{d}\mathbf x\hat{\Phi}^{\dagger}(\mathbf{x})\hat{\Phi}(\mathbf{x})$\footnote{It should be noted that $\hat{\Phi}^\dagger \hat{\Phi}$ can be interpreted as the particle density operator only when the commutator Eq.~(\ref{commutation}) holds. This fact is very important when we construct the coherent-state path integral formalism, in which the $\partial_t$-term is fixed by the commutator.}: $g=e^{i\alpha \hat{Q}}$ with a constant $\alpha\in\mathbb{R}$.  Through $\hat \Phi\rightarrow g^{-1}\hat\Phi g$, we end up with  a global phase shift on
$\hat{\Phi}(\mathbf{x})\rightarrow e^{i\alpha}\hat{\Phi}(\mathbf{x})$.
Moreover, one may employ the 1-form $U(1)$ symmetry according to a $U(1)$ gauge transformation
to describe 
a conventional electromagnetic field which can be minimally coupled to a matter field. 
Recently, the pursuit of more general
symmetry has been attributed to higher-rank symmetries and higher-moment
conservation, and the advancement brings
out plenty of interesting and exotic effects in various fields, which is 
not limited to the condensed matter. For the purpose of this paper, we
will fix our attention on quantum many-body systems that 
respect a HRS with higher-moment conservation, which, as we will see, imposes strong constraint on mobility of the constituent bosonic/fermionic particles. 
For convenience, \emph{hereafter}, we simply refer to these  systems as  many-fracton systems. 
Given a many-fracton system, one of the fundamental aspects is the appearance  of 
 SSB and  ODLRO \citep{YangODLRO} triggered by the Bose-Einstein condensation of fractons or subdimensional particles, which potentially leads to novel features beyond the conventional SSB.
The work \citet{2020PhRvR2b3267Y}   
studied the physics of  SSB and the associated ODLRO in many-fracton systems by introducing the notion of  fractonic superfluidity. In this context,  
  more general discussions  have   appeared, especially in the aspect of generalizing the celebrated  Mermin-Wagner theorem \citep{2020arXiv201003261C, 2022PhRvB.105o5107S,PhysRevB.106.245125}. 
Before focusing on the details of the SSB in fractonic superfluids, we
first describe the general formalism of HRS transformations in a many-fracton system.

The conservation of the higher moment and its relevance to fracton
order was pointed out in~\citet{Pretko2017,Pretko2018}. 
Above all, the symmetry perspective that have something to do with the conservation of the higher-moments have been extensively studied to be known as polynomial symmetries act on a real scalar field \citep{PhysRevD.79.064036,PhysRevD.88.101701,2014IJMPD..2343001H,2015CMaPh.340..985G}.
In general, one may consider a transformation that produces
a coordinate-dependent phase shift~\citep{Pretko2018,PhysRevX.9.031035}
\begin{equation}
\hat \Phi(\mathbf{x})\rightarrow\exp\left[i\sum_{a}\lambda_{a}P^{a}(\mathbf{x})\right]\hat \Phi(\mathbf{x})~,\label{eq:symm}
\end{equation}
where $\lambda_{a}$'s are   symmetry group parameters and $P^{a}(\mathbf{x})$
are some arbitrary  polynomials of coordinates. 
Given a system invariant under the transformation in Eq.~(\ref{eq:symm}),
a set of conserved charges follows
\begin{equation}
\mathcal{Q}_{a}=\int d^{d}\mathbf{x}\hat{\rho}(\mathbf x)P^{a}(\mathbf{x}) 
\end{equation}
with $\hat{\rho}(\mathbf x)=\hat \Phi^{\dagger}(\mathbf{x})\hat \Phi(\mathbf{x})$. We call such a symmetry a higher-rank
symmetry due to the resultant higher-rank gauge field upon gauging
\citep{Pretko2018}. 
One example is the conservation of dipole moments 
\begin{align}
Q_a=\int d^d \mathbf x  \hat \rho(\mathbf x)x^a \,\,(a=1,\cdots,d)
\end{align} which is exactly the center of mass of all particles. Hence in such a system, a single particle is a fracton with full localization in real space and can be thought with a divergent effective mass \citep{Pretko2018,2020PhRvR2b3267Y}. 
The transformations in Eq.~\eqref{eq:symm} commute with each other
and form a simple algebraic structure. However, these transformations 
no longer commutes with spatial translation and rotation symmetries. 
Instead, the higher-rank symmetry or
the polynomial shift symmetries extend the algebra
of spatial symmetries to a bigger multiple algebra \citep{PhysRevX.9.031035}.

\subsection{Multi-component bosons}\label{section_HRS_multiple}

If one considers a multi-component field whose 
  density operator $\boldsymbol{\rho}=(\hat{\rho}^{1},\hat{\rho}^{2},\cdots,\hat{\rho}^{d})$
can be regarded as a vector, one may define much more complicated higher-moment 
such as 
\begin{equation}
 Q=\int d^d\mathbf x\boldsymbol{\rho}\cdot\mathbf{x}, Q_{ab}=\int d^d\mathbf x \hat{\rho}_ax^b-\hat{\rho}_bx^a,\cdots,
 \end{equation} 
For an intuitive understanding \citep{2021PhRvR3d3176L}, we consider a classical picture at
two spatial dimensions with a two-component bosons with an angular moment $Q_{12}=\int d^d\mathbf x (\hat{\rho}_1x^2-\hat{\rho}_2x^1)$ conserved (see illustration in Fig.~\ref{Fig2D}). Suppose there are $N_{1}$($N_2$) particles of first(second) component particles. One may use Dirac function to express the $\hat{\rho}_{1}$
and $\rho_{2}$ in the eigenbasis, $\hat{\rho}_{a}=\sum_{i}^{N_{a}}\delta(\mathbf{x}-\mathbf{x}_{a,i})$,
which means that bosons of the 1st(2nd) components occupy
$\mathbf{x}_{a,i}$ with $i=1,2,\cdots,N_{a}$. Then $Q_{12}$ is
reduced to $Q_{12}=\sum_{i}^{N_{1}}x_{1,i}^{2}-\sum_{j}x_{2,j}^{1}$.
From this expression, one can conclude that if we move a 1st component
boson, in order to keep $Q_{12}$ invariant, the bosons is only allowed
to be movable in the 1st direction such that its coordinates $x_{1}^{2}$
is unchanged. Of course, one may collectively move bosons of both
components, such that the change in $\sum_{i}^{N_{1}}x_{1,i}^{2}$
can be canceled out by the change in $\sum_{i}^{N_{2}}x_{2,j}^{1}$
such that $Q_{12}$ is unchanged. Therefore, the conservation of $Q_{12}$ generates a lineon  in two spatial dimensions that is only allowed in one certain direction. Similarly, the conservation of angular moments $Q_{12},Q_{23}$ and $Q_{31}$ enforces a particle as a planeon that can only freely moves in a plane.
For the sake of convenience,  we still call all such many-body systems as many-fracton systems and the states due to condensation of lineons or planeons \citep{2020arXiv201003261C,2021PhRvR3d3176L,2022arXiv220108597Y,2022ChPhL..39e7101Y} are still said to be ``fractonic''.




\section{Minimal Hamiltonian, coherent state path integral, and spontaneous symmetry breaking}\label{section_minimal_model}

To begin with the realization of fracton physics in a concrete many-fracton model,
we emphasize that the standard commutation/anti-commutation relation shown in Eq.~(\ref{commutation})
allows us to define density operator in the form of   $\hat{\rho}(\mathbf x)=\hat\Phi^{\dagger}(\mathbf{x})\hat\Phi(\mathbf{x})$.  Thus,  one may expect a  time
derivative of the first order appears in the Lagrangian. More specifically, given a Hamiltonian $H[\hat{\Phi}(\mathbf{x})]$, within
the framework of coherent-state path integral \citep{Altland_simons_2010}, we can obtain the following  Lagrangian
\begin{equation}
\mathcal{\mathcal{L}}=i\phi^{*}\partial_{t}\phi-\mathcal{H}(\phi,\phi^{*})~,
\end{equation}
where $\phi(\mathbf{x},t)$ is the eigenvalue of $\hat{\Phi}\left(\mathbf{x}\right)$
on a coherent state $\hat{\Phi}\left(\mathbf{x}\right)|\phi\left(\mathbf{x},t\right)\rangle=\phi\left(\mathbf{x},t\right)|\phi\left(\mathbf{x},t\right)\rangle$. $\mathcal{H}(\phi,\phi^{*})$ is a Hamiltonian functional obtained by replacing $\Phi$ and $\Phi^\dagger$ in the Hamiltonian operator $H$ by eigenvalues $\phi$ and $\phi^*$ respectively (proper normal orderings are required). 
Therefore, one may legitimately interpret $\phi^{*}\phi$ as the particle
number density in the field-theoretical formalism. If the time derivative is of second order, the density operator is no longer written as $\Phi^\dagger\Phi$, which simultaneously changes the HRS generators. It should be noted
that a Wick rotation may be applied from imaginary time to real time,
which will benefit the treatment on the physics of zero temperature.

In this section, we will introduce minimal bosonic models that realize HRS and the associated SSB physics, i.e., fractonic superfluidity. Several properties of fracton superfluids are summarized in Table~\ref{Tab}.    Specifically, in Sec.~\ref{section_minimal_model_dipole}, we introduce the minimal model in which total dipoles are conserved, which was studied in \citet{2020PhRvR2b3267Y}. In Sec.~\ref{section_minimal_model_angular}, we introduce the minimal model in which total angular moments are conserved, which was studied in \citet{2020arXiv201003261C} and subsequently studied in \citet{2021PhRvR3d3176L,2022arXiv220108597Y,2022ChPhL..39e7101Y}. In Sec.~\ref{section_minimal_model_general}, we introduce general model construction. In Sec.~\ref{section_minimal_model_thm}, we introduce   generalization of Mermin-Wagner theorem studied in \citet{2022PhRvB.105o5107S,PhysRevB.106.245125}.

\subsection{Minimal model with dipole conservation}\label{section_minimal_model_dipole}

Below, we construct the minimal model with dipole conservation \citep{2020PhRvR2b3267Y}. 
In a conventional system, the leading hopping term is the
single particle hopping processes, e.g., $\nabla\hat{\Phi}^{\dagger}(\mathbf{x})\nabla\hat{\Phi}(\mathbf{x})$,
which obviously keeps a global $U(1)$ charge symmetry and does not
respect the higher-rank symmetry such as the dipole conservation. Instead,
a non-Gaussian Hamiltonian should come into play. 
Instead, we have to consider the cooperative motion with the two particles
propagating oppositely, giving rise to a kinetic term like 
\begin{equation}
H_0=\sum_{i,j}^{d}\!K_{ij}(\hat{\Phi}^{\dagger}\partial_{i}\partial_{j}\hat{\Phi}^{\dagger}-\partial_{i}\hat{\Phi}^{\dagger}\partial_{j}\hat{\Phi}^{\dagger})\!(\hat{\Phi}\partial_{i}\partial_{j}\hat{\Phi}-\partial_{i}\hat{\Phi}\partial_{j}\hat{\Phi}).
\end{equation}
For the purpose of SSB, one may introduce a Maxican-hat potential,
\begin{align}
V(\hat{\Phi}^{\dagger},\hat{\Phi})=-\mu\hat{\Phi}^{\dagger}\hat{\Phi}+\frac{g}{2}\hat{\Phi}^{\dagger}\hat{\Phi}^{\dagger}\hat{\Phi}\hat{\Phi}\,,\label{eq_mex}
\end{align}
where $\mu$ is the chemical potential and $g>0$ describes onsite
repulsive interaction. Most importantly, the system has the first
order of the the time derivative. The Euler-Lagrange equation (EL)
should be complicated due to the higher-order derivative. For example,
specific to the minimal model of dipole conservation symmetry, the
EL equation is 
\begin{equation}
\partial_{t}\frac{\delta\mathcal{L}}{\delta\partial_{t}\phi}=\frac{\delta\mathcal{L}}{\delta\phi}-\sum_{i}^{d}\partial_{i}\frac{\delta\mathcal{L}}{\delta\partial_{i}\phi}+\sum_{i,j}^{d}\partial_{i}\partial_{j}\frac{\delta\mathcal{L}}{\delta\partial_{i}\partial_{j}\phi}.
\end{equation}
The third term results from the cooperative two-particle hopping processes and  is usually absent in a conventional system introduced in e.g., textbook of quantum field theory and classical mechanics. 
The resulting EL equation or the non-linear Schroedinger equation
has complicated non-Gaussian kinetic terms,
\begin{align}i\partial_{t}\phi  =& \sum_{i,j}^{d}K_{ij}\partial_{i}\partial_{j}\left[\phi^{\ast}\left(\phi\partial_{i}\partial_{j}\phi-\partial_{i}\phi\partial_{j}\phi\right)\right]\notag\\
 & +2K_{ij}\partial_{i}\left[\partial_{j}\phi^{\ast}\left(\phi\partial_{i}\partial_{j}\phi-\partial_{i}\phi\partial_{j}\phi\right)\right]\notag\\
 & +K_{ij}\partial_{i}\partial_{j}\phi^{\ast}\left(\phi\partial_{i}\partial_{j}\phi-\partial_{i}\phi\partial_{j}\phi\right)\notag \\
 & -\mu\phi+g\rho\phi\,.
\end{align}
One may verify that a plane-wave ansatz $\phi\propto\exp\left(i\omega t-i\mathbf{k}\cdot\mathbf{x}\right)$
is no longer mobile with a flat dispersion $\omega=0$, which indeed
demonstrates the fundamental particle $\hat\Phi$ is a fracton with fully
restricted mobility. Moreover, from the Noether theorem, we can have
two types of conserved currents, $J_{i}$ and $\mathcal{D}_{i}^{\left(a\right)}$
respectively corresponding to the global U(1) charge and the higher-rank
U(1) charges, 
\begin{align}
Q  =&\int\!\mathrm{d}^{d}\mathbf{x}\,\phi^{\ast}\phi\,,\\
J_{i}  =&i\sum_{j}^{d}K_{ij}\partial_{j}\left[\phi^{\ast2}\left(\phi\partial_{i}\partial_{j}\phi-\partial_{i}\phi\partial_{j}\phi\right)-\text{c.c.}\right]\,,\\
Q^{\left(a\right)}  =&\int\!\mathrm{d}^{d}\mathbf{x}\,x^{a}\phi^{\ast}\phi\,,\\
\mathcal{D}_{i}^{\left(a\right)}  =&i\sum_{j}^{d}K_{ij}x^{a}\partial_{j}\left[\phi^{\ast2}\left(\phi\partial_{i}\partial_{j}\phi-\partial_{i}\phi\partial_{j}\phi\right)-\text{c.c.}\right]\nonumber \\
 & -i\sum_{j}^{d}K_{ij}\delta_{a}^{j}\left[\phi^{\ast2}\left(\phi\partial_{i}\partial_{j}\phi-\partial_{i}\phi\partial_{j}\phi\right)-\text{c.c.}\right]\,.
\end{align}
In fact, one may isolate the many-body currents \citep{2020PhRvR2b3267Y}
\begin{equation}
\Xi_{ia}=x^{a}J_{i}-\mathcal{D}_{i}^{(a)}
\end{equation}
which is symmetric about its indices. $\Xi_{ia}$ has a nice relation to charge current 
\begin{equation}
J_{i}=\sum_{a=1}^{d}\partial_{a}\Xi_{ia}
\end{equation}
which implies the higher-rank symmetry here requires the each component of $J_i$
be a total derivative.
The $\Xi_{ia}$ represents
a pure effect during many-body hopping processes and satisfies a generalized
conservation law 
\begin{equation}
\partial_{t}\rho+\sum_{i,a}\partial_{i}\partial_{a}\Xi_{ia}=0,
\label{cons_law}
\end{equation}
while the term $x^{a}J_{i}$ describes motions of each single particle
$\mathbf{x}$ with currents $J_{i}$. Recently, the conservation law in Eq.~\eqref{cons_law} and its variants from various theoretical considerations
have played a crucial role in physics of fracton systems and higher moment conservations, see, e.g., \citet{Pretko2018,Pretko2017a,2020PhRvL.125x5303F,2020PhRvB.101h5106S,Seiberg2019arXiv1909,2022ScPP1250D,2020PhRvR2c3124G}.

An ODLRO can operate by following the routine towards a conventional
superfluid phase with a Mexican-hat potential 
\begin{equation}
V(\phi)=-\mu\rho+\frac{g}{2}\rho^{2}~,
\end{equation}
which is path-integral representation of Eq.~(\ref{eq_mex}).  Explicitly, when $\mu<0$, we have a normal phase. While a positive
chemical potential $\mu>0$ leads to the minimal value of $V(\phi)$
at $\left\vert \phi\right\vert =\sqrt{\rho_{0}}\equiv\sqrt{\frac{\mu}{g}}$,
then  the ground state manifold support large degeneracies parametrized by $\{\theta_0,\beta_1,\beta_{2},\cdots,\beta_d\}$:  
\begin{equation}
|\text{GS}_{\beta_{i}}^{\theta_{0}}\rangle=\bigotimes_{\mathbf{x}}|\text{GS}_{\beta_{i}}^{\theta_{0}}\rangle_{\mathbf{x}}\,,
\end{equation}
where $|\text{GS}_{\beta_{i}}^{\theta_{0}}\rangle_{\mathbf{x}}$
describes particles at position $\mathbf{x}$ with 
\begin{equation}
|\text{GS}_{\beta_{i}}^{\theta_{0}}\rangle_{\mathbf{x}}=\frac{1}{C}\exp\left[\sqrt{\rho_{0}}e^{i\left(\theta_{0}+\sum_{i}^{d}\beta_{i}x^{i}\right)}\Phi^{\dagger}(\mathbf{x})\right]|0\rangle~,
\end{equation}
with $C=e^{\frac{1}{2}\rho_{0}}$ being the normalization factor. The
ground state $|\text{GS}_{\beta_{i}}^{\theta_{0}}\rangle$ comprises
equal-weight superposition over all possible numbers of particles
that is modulated by a phase factor and it characterizes condensation
of a macroscopically large number of particles at a state with momentum
$\mathbf{k}=(\beta_{1},\cdots,\beta_{d})$ by observing $|\mathrm{GS}_{\beta_{i}}^{\theta_{0}}\rangle=\exp[\sqrt{\rho_{0}}e^{i\theta_{0}}\hat{\Phi}^{\dagger}(\mathbf{k})]|0\rangle$
where $\hat{\Phi}^{\dagger}(\mathbf{k})$ is the Fourier transformation
of $\hat{\Phi}^{\dag}(\mathbf{x})$. The most significant feature
here is the formation of a (classical) ODLRO with a nonvanishing spatially varying order parameter defined
by 
\begin{equation}
\langle\hat{\Phi}(\mathbf{x})\rangle=\langle\text{GS}_{\beta_{i}}^{\theta_{0}}|\hat{\Phi}(\mathbf{x})|\text{GS}_{\beta_{i}}^{\theta_{0}}\rangle=\sqrt{\rho_{0}}e^{i(\theta_{0}+\sum_{i}\beta_{i}x^{i})}.
\end{equation}
Instead of a charge current, the many-body current $\mathbf{\Xi}$
takes the role of the supercurrent of the fractonic superfluid phase
and above the critical value 
\begin{equation}
(\Xi_{s})_{\mathrm{max}}=\frac{3\sqrt{3\kappa}\mu^{2}}{16\sqrt{g^{3}}},
\end{equation}
the fractonic superfluid phase will no longer survive.

\begin{table}[t]
\caption{Comparison between two-point correlators of conventional superfluid
and \textit{isotropic} fractonic superfluid at zero temperature after
quantum fluctuations are included. $d$ is spatial dimension. The
effective theories are given by Eqs.~(\protect\ref{eq:corr_iso})
and (\protect\ref{eq:corr_con}) respectively. The former has $c=\sqrt{\kappa g\rho_{0}}$
and coherence length $\xi_{\mathrm{coh}}=2\pi\sqrt{\kappa/(4\rho_{0}g)}$
while the latter with isotropic coupling constant $K_{ij}=\frac{1}{2}\kappa$
has $c=\sqrt{\kappa g\rho_{0}^{2}}$ and coherence length $\xi_{\mathrm{coh}}=2\pi\sqrt[4]{\kappa/4g}$.
A many-fracton system is fully disordered (marked by $\times$ ) in
$d=1$ and algebraically ordered (AO) in $d=2$. It has a stable ODLRO,
i.e., a true superfluid (marked by \checkmark), when $d\protect\geq3$
at zero temperature. Here $\gamma$ is the Euler's constant. 
From \citet{2020PhRvR2b3267Y}.
}
\label{Tab:Correlation}
\setlength{\tabcolsep}{7mm}{
   \begin{tabular}{ccccc}
\specialrule{0.1em}{1pt}{0pt}
 $d$  & \multicolumn{2}{c}{{Conventional system}} & \multicolumn{2}{c}{{Many-fracton system}} \tabularnewline
\specialrule{0.05em}{0.5pt}{0.5pt} $1$  & $\rho_{0}e^{-\frac{\gamma g}{4\pi c}}(\pi r/\xi_{\mathrm{coh}})^{-\frac{g}{2\pi c}}$  & AO  & $\rho_{0}e^{-\frac{g}{2c}(\pi r-\xi_{\mathrm{coh}}/\pi^{\frac{3}{2}})}$  & $\times$ \tabularnewline
$2$  & $\rho_{0}e^{-\frac{g}{2\pi c}\xi_{\mathrm{coh}}^{-1}}$  & \checkmark  & $\rho_{0}e^{-\frac{\gamma g}{4\pi c}}(r/\xi_{\mathrm{coh}})^{-\frac{g}{2\pi c}}$  & AO \tabularnewline
$\geq3$  & $\rho_{0}e^{-g\frac{\pi^{\frac{d-3}{2}}}{2\left(d-1\right)c}\xi_{\mathrm{coh}}^{1-d}}$  & \checkmark  & $\rho_{0}e^{-\frac{g}{c}\frac{\pi^{\frac{d}{2}-2}}{2(d-2)}\xi_{\mathrm{coh}}^{2-d}}$  & \checkmark \tabularnewline
\specialrule{0.1em}{1pt}{0pt}  &  &  &  & \tabularnewline
\end{tabular}
}
\end{table}

The stability
of the ground state should rely on the property of the Goldstone modes,
and its effective theory can be reached by making an expansion around
the classical field configuration. In the long wavelength limit, we
obtain an effective Lagrangian for field $\theta$ to the leading
order, 
\begin{equation}
\mathcal{L}=\frac{1}{2g}(\partial_{t}\theta)^{2}-\rho_{0}^{2}\sum_{i,j}^{d}K_{ij}(\partial_{i}\partial_{j}\theta)^{2}~,
\end{equation}
which represents the Gaussian fluctuations. Here we only have one
single Goldstone mode, even through the global U(1) and dipole U(1)
get broken simultaneously. Instead, the Goldstone mode picks up a
higher-dispersion relation. The effect of quantum fluctuations can
be deduced by calculating the two-point correlator $C\!\left(\mathbf{x}\right)=\langle\hat{\Phi}(\mathbf{x})\hat{\Phi}{}^{\dagger}(\mathbf{x})\rangle$.
For the isotropic case $K_{ij}=\frac{1}{2}\kappa$ for any $i,j$,
the Goldstone mode $\theta$ has a quadratic dispersion relation $\omega=\sqrt{\kappa g\rho_{0}^{2}}\left\vert \mathbf{k}\right\vert ^{2}\equiv c\left\vert \mathbf{k}\right\vert ^{2}$
that recovers rotational symmetry, and effective Lagrangian for the
Goldstone mode takes the form as 
\begin{equation}
\mathcal{L}=\frac{1}{2g}(\partial_{t}\theta)^{2}-\frac{1}{2g}c^{2}\left(\nabla^{2}\theta\right)^{2}\label{eq:corr_iso}~.
\end{equation}
It possesses an emergent Lifshitz spacetime symmetry 
~\citep{Hoava2009,Hoava2009a,Xu2010,PhysRevB.106.045112,2022arXiv220910030G}. 
 By looking at the asymptotic behavior of $C\left(\mathbf{x}\right)=\rho_{0}e^{-\frac{1}{2}\langle(\theta(\mathbf{x})-\theta(\mathbf{0}))^{2}\rangle}$ (see also Table~\ref{Tab:Correlation} for comparison)
\begin{equation}
C(\mathbf{x})=\begin{cases}
\rho_{0}e^{-\frac{g}{2c}(\pi r-\pi^{-\frac{3}{2}}\xi_{\mathrm{coh}})} & d=1\\
\rho_{0}e^{-\frac{\gamma g}{4\pi c}}\left(\frac{r}{\xi_{\mathrm{coh}}}\right)^{-\frac{g}{2\pi c}} & d=2\\
\rho_{0}e^{-\frac{g}{c}\frac{\pi(\frac{d}{2}-2)}{2(d-2)}}\xi_{\mathrm{coh}}^{2-d} & d\geq3
\end{cases}
\end{equation}
with $\xi_{\mathrm{coh}}$ being the coherent length, we can find
that only when our spatial dimension $d>2$ does a superfluid survive
quantum fluctuations. In particular, the correlator $C(\mathbf{x})$
approaches zero in dimension $d=1$ and $2$ in a large distance.
In $d=2$, $C(\mathbf{x})$ decays in a power-law pattern,
which is similar to a conventional superfluid in $d=1$. Another aspect
of the higher-order dispersion is specific heat capacity $c_{v}$
that is proportional to $T^{\frac{d}{2}}$ in $d$ space dimensions,
while the specific heat capacity is physical meaningless when $d>3$.
As a comparison, for a conventional superfluid with Lagrangian $\mathcal{L}=i\phi^{\ast}\partial_{t}\phi-\frac{1}{2}\kappa\left\vert \nabla\phi\right\vert ^{2}-V\left(\phi\right)$
with $V\left(\phi\right)$, the effective theory for the Goldstone
mode is 
\begin{equation}
\mathcal{L}=\frac{1}{2g}(\partial_{t}\theta)^{2}-\frac{1}{2g}c^{2}\left(\nabla\theta\right)^{2} ~,\label{eq:corr_con}
\end{equation}
where Goldstone mode has a linear dispersion relation $\omega=\sqrt{kg\rho_{0}}\left\vert \mathbf{k}\right\vert \equiv c\left\vert \mathbf{k}\right\vert $
and quantum fluctuation will kill a superfluid phase in one spatial
dimension at zero temperature. Note that $\theta$ is an angular-valued
field and is defined $\mathrm{mod}2\pi$, the defect excitations will
be clarified in Sec. \ref{section_kt}.


\subsection{Minimal model with angular moment conservation\label{section_minimal_model_angular}}

\begin{figure}[t]
\centering 
\includegraphics[scale=0.05]{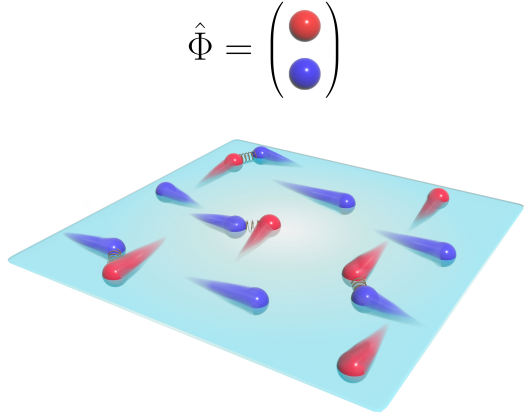}
\caption{Illustration of the interacting system in two dimensions. The red and blue balls respectively represent two components $a=1,2$, which move along distinct orthogonal directions. The spring between two balls 
represents the interaction due to the $K$-term in Eq.~\eqref{eq:kab_angular}. From~\citet{2020arXiv201003261C}.
}
\label{Fig2D}
\end{figure}

The second model involves a system comprising many lineons that are
subject to the angular moment conservation to restrict its propagation
only along one certain spatial direction \citep{2020arXiv201003261C}.
The prototypical model Hamiltonian contains a $d$-component fields
$\hat{\Phi}_{a}$ that obey a standard commutation relation. To conserve
the angular moment $Q_{ab}=\int d^{d}\mathbf{x}(\hat{\rho}_{a}x^{b}-\hat{\rho}_{b}x^{a})$,
the minimal hopping terms are $H_0= \sum_{a=1}^{d}\partial_{a}\hat{\Phi}_{a}^{\dagger}\partial_{a}\hat{\Phi}_{a}$
and 
\begin{equation}
\!\!\!H_2=\sum_{a\neq b}^{d}K_{ab}(\hat{\Phi}_{a}^{\dagger}\partial_{a}\hat{\Phi}_{b}^{\dagger}+\hat{\Phi}_{b}^{\dagger}\partial_{a}\hat{\Phi}_{a}^{\dagger})(\hat{\Phi}_{a}\partial_{a}\hat{\Phi}_{b}+\hat{\Phi}_{b}\partial_{b}\hat{\Phi}_{a})~.\label{eq:kab_angular}
\end{equation}
The Hamiltonian $H= H_0+H_2$  is invariant under the transformation $(\hat{\Phi}_{a},\hat{\Phi}_{b})\rightarrow(\hat{\Phi}_{a}e^{i\lambda_{ab}x^{b}},\hat{\Phi}_{b}e^{-i\lambda_{ab}x^{a}})$
with $\frac{d(d-1)}{2}$ anti-symmetric real parameter $\lambda_{ab}=-\lambda_{ba}$.
The parameters $\lambda_{ab}$ have the dimension of $[x]^{-1}$ and
quantization of the related charges is expected to coincide with a
momentum in a periodic boundary condition. This symmetry involves
local coordinate $\mathbf{x}=(x^{1},x^{2},\cdots,x^{d})$ and imposes
that strong constraints on particles' propagation. In $2$D, conservation
of $Q_{12}=\int d^{2}x\left(\hat{\rho}_{1}x^{2}-\hat{\rho}_{2}x^{1}\right)$
requires the velocity shall be parallel to a vector $(\hat{\rho}_{1},\hat{\rho}_{2})$.
In three spatial dimensions, we have $3$ angular charge moments $Q_{12},Q_{23},Q_{13}$,
such that a particle only propagates in the direction parallel to
$\left(\hat{\rho}_{1},\hat{\rho}_{2},\hat{\rho}_{3}\right)$ (for illustration of 2D case, see Fig.~\ref{Fig2D}). Generally,
the fundamental particles in $d$ spatial dimensions move with velocity parallel to
$\left(\hat{\rho}_{1},\cdots,\hat{\rho}_{d}\right)$. 

\begin{figure}[t]
    \centering
    \includegraphics[width=0.6\columnwidth]{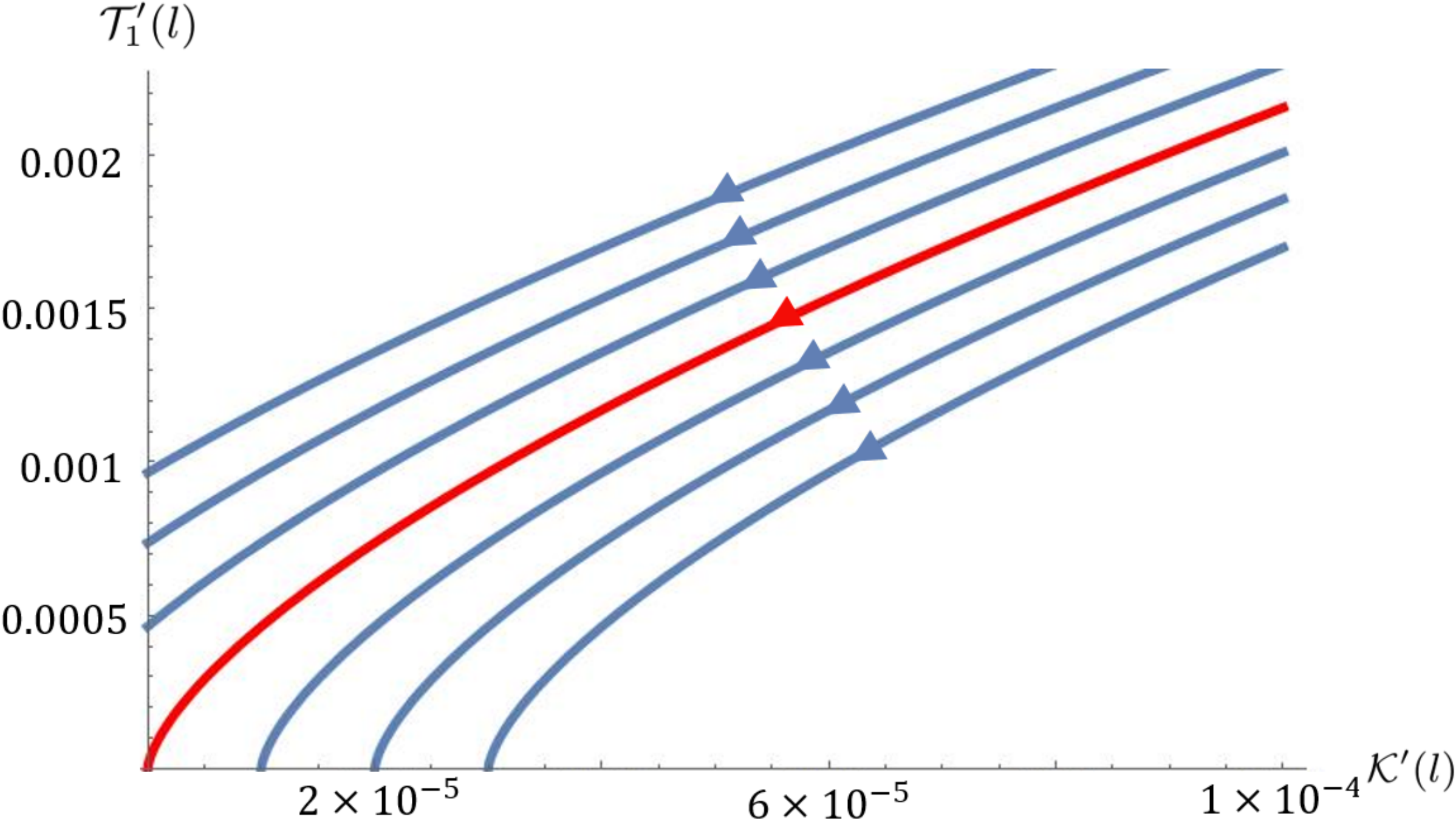}\\
    \caption{RG flow for $\mathcal{T}_{1}'(l)$ and $\mathcal{K}'(l)$ in two spatial dimensions. Here the red line is a separatrix below which higher-rank symmetry emerges. The quantity $\mathcal{T}_{1}'(l)$ is related to the `off-diagonal' kinetic energy $\sum_{a\neq b }^d\frac{t_{ab}}{2}(\partial_b\phi_a^*)(\partial_b\phi_a)$ and $\mathcal{K}'(l)$
    corresponds to $K_{ab}$ in Eq.~\eqref{eq:kab_angular}.
    From~\citet{2021PhRvR3d3176L}. 
     }
    \label{figure_RGflow}
\end{figure}

While higher-rank symmetric microscopic models may seem unrealistic, highly anisotropic, and fine-tuned, 
it has been found that in many condensed matter systems, 
symmetry is significantly enhanced at low energies. 
For instance, graphene microscopically built by non-relativistic electrons exhibits the emergence of Lorentz symmetry. 
This leads us to question whether higher moments could be conserved and higher-rank symmetry could emerge as a long-wavelength, low-energy limit. 
This concept has been explored in recent research \citep{2021PhRvR3d3176L}, 
with the perspective of experimental realization in mind. One possible approach is to consider the coexistence of the usual kinetic terms, which break angular moment conservation, and the terms in Eq.~(\ref{eq:kab_angular}), 
followed by the application of the traditional renormalization group method. 
It has been shown that, by including the one-loop correction in two spatial dimensions, 
as illustrated in Fig.~\ref{figure_RGflow}, the RG flows to a region in which higher-rank symmetry emerges. 
Consequently, this emergent higher-rank symmetry possesses the advantage of being robust against symmetry-breaking perturbations. 
Thus, this scenario holds promise for more flexible realization of exotic higher-rank symmetry and higher-moment conservation in both theoretical and experimental studies.

From the Neother theorem, we can have two types of Neother charges and currents. The $Q_{a}$ with charge density $\rho_{a}$ and currents $J_{i}^{a}$ are related to the global $U(1)$
 read,
\begin{align}
Q^{a}  =&\int \mathrm d^{d}\mathbf{x}\phi _{a}^{\ast }\phi _{a}\equiv \int \mathrm d^{d}%
\mathbf{x}\rho_{a}  \label{u(1)charge} \,,\\
J_{i}^{a}   =&  iK _{ai}\rho _{a}\left(\phi _{i}\partial _{a}\phi _{i}^{\ast }- \phi _{i}^{\ast }\partial _{a}\phi
_{i}\right) \nonumber\\
&+iK _{ai}\rho
_{i}\left(\phi _{a}\partial
_{i}\phi _{a}^{\ast }- \phi _{a}^{\ast }\partial _{i}\phi _{a}\right)  \notag \\
&+i\left(\phi _{i}\partial
_{i}\phi _{i}^{\ast }- \phi _{i}^{\ast }\partial _{i}\phi _{i}\right) \delta _{ai}~,  \label{u(1)current}
\end{align}%
which satisfies the continuity equations ($a=1,2,\cdots,d$) 
\begin{align}
\partial _{t}\rho
^{a}+\sum_{i}\partial _{i}J_{i}^{a}=0.
\end{align} 
For the angular moments $Q_{ab}$ (with density $\rho_{ab}$) and currents $D_i^{ab}$, we have
\begin{align}
Q_{ab} &=\int \mathrm d^{d}x\left( \rho _{a}x^{b}-\rho_{b}x^{a}\right) \equiv \int \mathrm d^{d} x \rho_{ab} \\
D_{i}^{ab} &=x^{b}J_{i}^{a}-x^{a}J_{i}^{b}  \label{dipolecurrent}
\end{align}%
with $\rho_{a}$ and $J_{i}^{a}$ as $U(1) $ charge and current
in Eqs.~(\ref{u(1)charge}) and (\ref{u(1)current}).  The continuity equation
\begin{align}
\partial _{t}\rho^{ab}+\sum_{i=1}^{d}\partial _{i}D_{i}^{ab}=0
\end{align}
is automatically satisfied as long as the currents $J_{b}^{a}$
obey the relations $J_{b}^{a}=J_{a}^{b}$.

Upon $\mu>0$ in the potential $V=\sum_{a}-\mu\hat{\Phi}_{a}^{\dag}\hat{\Phi}_{a}+\frac{g}{2}\hat{\Phi}_{a}^{\dag}\hat{\Phi}_{a}^{\dag}\hat{\Phi}_{a}\hat{\Phi}_{a}$,
one may obtain a fractonic superfluid $d\mathsf{SF}^{1}$ with lineon
condensation. The ground state then can be described by 
\begin{equation}
\vert\text{GS}_{\theta_{a}}^{\beta_{ab}}\rangle=\prod\limits _{a=1}^{d}\exp[\sqrt{\rho_{0}}e^{i\left(\theta_{a}+\sum_{b=1}^{d}\beta_{ab}x^{b}\right)}\hat{\Phi}_{a}^{\dagger}\left(\mathbf{x}\right)]|0\rangle~,
\end{equation}
with real parameters $\theta_{a}$ and $\beta_{ab}$ $(\beta_{ab}=-\beta_{ba},a,b=1,\cdots d)$.
 The ground states $\vert\text{GS}_{\theta_{a}}^{\beta_{ab}}\rangle$
feature (classical) ODLRO with a nonvanishing spatially varying  order parameter $\hat{\Phi}_{a}(\mathbf{x})$,
\begin{equation}
\langle\text{GS}_{\theta_{a}}^{\beta_{ab}}\vert\hat{\Phi}_{a}(\mathbf{x})\vert\text{GS}_{\theta_{a}}^{\beta_{ab}}\rangle=\sqrt{\rho_{0}}\exp(i\theta_{a}+i\sum_{b}\beta_{ab}x^{b}),
\end{equation}
which denotes a fractonic superfluid phase with lineon condensation.
The expectation value oscillates as a plane-wave with fixed momentum
$\mathbf{k}_{a}=\left(\beta_{a1},\beta_{a2},\cdots,\beta_{ad}\right)$
for the $a$-component particle. In this sense, we can rewrite $|\text{GS}_{\theta_{a}}^{\beta_{ab}}\rangle=\prod\limits _{a=1}^{d}\exp\left[\sqrt{\rho_{0}}e^{i\theta_{a}}\hat{\Phi}_{a}^{\dagger}(\mathbf{k}_{a})\right]|0\rangle$
with $\hat{\Phi}_{a}^{\dagger}\left(\mathbf{k}_{a}\right)$ being
the Fourier transformation of $\hat{\Phi}_{a}^{\dagger}(\mathbf{x})$.
These features arise from restricted mobility of condensed particles.
As a side note, the ground state, which carries finite momentum, looks
like a Fulde-Ferrell-Larkin-Ovchinnikov (FFLO) state \citep{PhysRev135A550,larkin:1964zz}.
The effective theory for the Goldstone modes then becomes 
\begin{equation}
\mathcal{L}=\sum_{a}\frac{1}{2g}\left(\partial_{t}\theta_{a}\right)^{2}-\rho_{0}\left(\partial_{a}\theta_{a}\right)^{2}-\sum_{a,b}\frac{1}{2}K_{ab}\rho_{0}^{2}\left(\partial_{a}\theta_{b}+\partial_{b}\theta_{a}\right)^{2}.
\label{eq:eff2sf1}
\end{equation}
It stays invariant under the transformation $\theta_{a}\rightarrow\theta_{a}+\lambda_{a}+\sum_{b}\lambda_{ab}x^{b}$
with $\lambda_{ab}=-\lambda_{ba}$. There are $d$ branches of gapless
Goldstone modes with entangled motions arising from the $K_{ab}$ terms.
One may concentrate oneself on the two-dimensional case by introducing
the canonical modes $\Theta_{\pm}(\mathbf{k})$ by the Bogoliubov
transformation $\Theta_{+}(\mathbf{k})=\cos\frac{\varphi_{\mathbf{k}}}{2}\theta_{1}(\mathbf{k})+\sin\frac{\varphi_{\mathbf{k}}}{2}\theta_{2}(\mathbf{k})$,
$\Theta_{-}(\mathbf{k})=-\sin\frac{\varphi_{\mathbf{k}}}{2}\theta_{1}(\mathbf{k})+\cos\frac{\varphi_{\mathbf{k}}}{2}\theta_{2}(\mathbf{k})$
with $\tanh\frac{\varphi_{\mathbf{k}}}{2}=\frac{2K\rho_{0}k_{1}k_{2}}{\Delta(\mathbf{k})^{2}}$.
The modes 
$
\Theta_{\pm}(\mathbf{k})$ have dispersion relations 
\begin{equation}
\epsilon_{\pm}(\mathbf{k})=\sqrt{g\rho_{0}\left[(1+K\rho_{0})\mathbf{k}^{2}\pm\Delta(\mathbf{k})\right]}
\end{equation}
with $\Delta(\mathbf{k})=\sqrt{(k_{1}^{2}-k_{2}^{2})^{2}(K\rho_{0}-1)^{2}+4K^{2}\rho_{0}^{2}k_{1}^{2}k_{2}^{2}}$.
The stability of the $d\mathsf{SF}^{1}$ depends on the long-distance
behavior of the order parameter correlator under the influence of
quantum fluctuations. 
\begin{align}
 & \langle\text{GS}_{\theta_{a}}^{\beta_{ab}}\vert\Phi_{a}^{\dagger}\left(\mathbf{x}\right)\Phi_{b}\left(\mathbf{0}\right)\vert\text{GS}_{\theta_{a}}^{\beta_{ab}}\rangle\nonumber \\
= & \rho_{0}\exp[i\sum_{c}(\beta_{bc}-\beta_{ac})x^{c}]\langle e^{-i\Theta_{a}(\mathbf{x})}e^{i\Theta_{b}(\mathbf{0})}\rangle~.
\end{align}
Owing to the linear tendency at small momentum $\mathbf{k}$, the correlator
$\left\langle \Theta_{\pm}\left(\mathbf{x}\right)\Theta_{\pm}(\mathbf{0})\right\rangle \sim\frac{1}{2\pi|\mathbf{x}|}$,
giving rise to a long distance behavior,
with $\left\langle \Phi_{a}^{\dagger}\left(\mathbf{x}\right)\Phi_{b}\left(\mathbf{0}\right)\right\rangle =\rho_{0}\exp[i\sum_{c}(\beta_{bc}-\beta_{ac})x^{c}]$ at $\vert\mathbf{x}\vert\rightarrow\infty$
as a finite value modulated by a plane wave. It confirms a true long-range
order $2\mathsf{SF}^{1}$ that survives against quantum fluctuations
when the lineons condensate simultaneously in zero temperature. Since
the quantum fluctuations are weaker in higher-dimensions, a fractonic
superfluid phase $d\mathsf{SF}^{1}$ stays stable in two spatial dimensions
$d=2$ and higher $d>2$.

In fact, one may apply a Landau criteria for stability of $d\mathsf{SF}^1$ \citep{2022ChPhL..39e7101Y}.
In $d\mathsf{SF}^{1}$, the current $J_{\text{m}}=\sum_{a}\left|J_{aa}\right|/d $
 is along the movable direction and the current  $J_{\text{im}}=\sqrt{\sum_{i}\sum_{a(\neq i)}\left|J_{ai}\right|^{2}/d}$
   is along the immovable direction. These two currents have different critical values. In the movable direction, the critical current is  $J_{\text{m}}^{\text{max}}  =\frac{4\mu^{3/2}}{3\sqrt{3}g}\label{eq:m critical current}$, while at 
 the immovable direction, the critical current is $J_{\text{im}}^{\text{max}}  =\frac{3\sqrt{3K}\mu^{2}}{8g^{3/2}}$. 
Once the currents exceeds the critical values, the superfluid density vanishes
and the system is no longer in the superfluid phase.

\subsection{General Hamiltonian for single-component many-fracton models}\label{section_minimal_model_general}

One may generalize many-fracton model into a large class \citep{2020Fracton,2020arXiv201003261C,2020PhRvR...2d3219W}. 
As far as the single-component many-fraton model, one can see that 
the kinetic
term can be formulated as 
 \begin{align}
\mathcal{H}_{N}= & \sum_{i_{1}i_{2}\cdots i_{N+1}}^{d}K_{i_{1}i_{2}\cdots i_{N+1}}\left(\hat{\Phi}^{\dag}\right)^{N+1} 
\left(\nabla_{i_{1}i_{2}\cdots i_{N+1}}^ {}\log\hat{\Phi}^{\dag}\right) \notag \\ 
&\cdot \hat{\Phi}^{N+1}\left(\nabla_{i_{1}i_{2}\cdots i_{N+1}}^{}\log\hat{\Phi}\right)~,
\end{align}
with $\nabla_{i_{1}i_{2}\cdots i_{n}}^ {}=\partial_{i_{1}}\partial_{i_{2}}\cdots\partial_{i_{n}}$
and the summation for each index is over all spatial dimensions.  
Here the canonical communication in Eq.~\eqref{commutation} is assumed to ensure 
a first order time derivative.
The coupling constant $K_{i_{1}i_{2}\cdots i_{N+1}}$ can be anisotropic
and it is fully symmetric with its indexes. When $N=0$, it reduces to
a Gaussian theory and when $N=1$, it reduces to the many-fracton
model that preserves rank-1 U(1) symmetry. 
Although $\log\hat\Phi$ is a multivalued function and has singularity,
the kinetic term turns out to be well-defined. The system is invariant
under a transformation $\hat \Phi\rightarrow\exp\left(i\delta\theta\right)\hat\Phi$
where $\delta\theta$ is polynomials of degree $N$ of local coordinates
$\delta\theta\!=\!\sum_{i_{1}\cdots i_{N}}\!\mathcal{D}_{i_{1}i_{2}\cdots i_{N}}x^{i_{1}}x^{i_{2}}\cdots x^{i_{N}}+\!\cdots\!+\sum_{i}\mathcal{D}_{i}x^{i}+\mathcal{D}$
where $\mathcal{D}_{i_{1}\cdots i_{l}}$ is a symmetric real tensor
of rank-$l$ with respect to spatial indexes. And the related conserved
charges have the form as $Q^{\left(C\left(x^{a}\right)\right)}=\int d^{d}\mathbf{x}\rho C\left(x^{a}\right)$
with $C\left(x^{a}\right)$ being a homogeneous polynomials with degree-$p$
and $p\leq N$.

Take an isotropic coupling constant $\mathcal{D}_{i_{1}i_{2}\cdots i_{N+1}}=\frac{1}{2}\kappa$
as an example. If we take a Mexican-hat potential chemical potential
$\mu>0$, we have degenerate vacuum with finite uniform density distribution
$\rho=\rho_{0}$. Through the same processes, we can derive an effective
theory for the quantum fluctuation field $\theta$ after condensation
\begin{equation}
\mathcal{L}=\frac{1}{2g}\left(\partial_{t}\theta\right)^{2}-\frac{1}{2g}c^{2}\left(\nabla^{N+1}\theta\right)^{2}.\label{eq:Glodstone_gn}
\end{equation}
The effective theory describes Goldstone mode $\theta$ with a dispersion
relation $\omega=\sqrt{\kappa g\rho_{0}^{N+1}}\left\vert \mathbf{k}\right\vert ^{N+1}\equiv c\left\vert \mathbf{k}\right\vert ^{N+1}$.
The calculation on the correlator $C\left(\mathbf{x}\right)$ 
shows that $C\left(\mathbf{x}\right)$ decays to zero when spatial
dimension is lower than $d<N+2$ at zero temperature. In particular,
$C\left(\mathbf{x}\right)$ decays in a power-law pattern at zero
temperature at spatial dimension $d=N+1$.

\subsection{General development on Mermin-Wagner Theorem}\label{section_minimal_model_thm}

 From the above two minimal models (see \citet{2020PhRvR2b3267Y,2020arXiv201003261C} as well as \citet{2021PhRvR3d3176L,2022arXiv220108597Y,2022ChPhL..39e7101Y}), one may tell that dipole symmetry cannot be spontaneously broken at the spatial dimension $d\leq 2$ at $T=0$ and  the angular momentum conservation symmetry cannot be spontaneously broken at spatial dimension $d\leq 1$ at $T=0$ (see Table~\ref{Tab:Correlation}). 
Keeping  these established facts from concrete model studies, it is natural to move on to a discussion of the Mermin-Wagner theorem for a general higher-rank symmetry. 
The conventional Mermin-Wagner theorem states that the continuous symmetries cannot be 
spontaneously broken at finite temperature in the systems with sufficiently short-range 
interactions in the spatial dimensions $d\leq 2$. 
The coordinate-dependence property of a higher-rank symmetry will influence the dispersion of the Golstone modes, which possibly leads to  more severe quantum destructive interference in the path-integral.
The generalization shall clarify if and when
higher-rank symmetries can be spontaneously broken, both in thermal
equilibrium and at zero temperature. For example, in \citet{2022PhRvB.105o5107S}  they generalized
Mermin-Wagner theorem to the case where an arbitrary maximal multipole
group is spontaneously decomposed into the trivial subgroup with an
effective theory for the Goldstone mode as in Eq.~(\ref{eq:Glodstone_gn}),
the SSB cannot be spontaneously broken for $d\leq2(N+1)$ at finite temperature. 
They also discussed the generalized Imry-Ma arguements for the
robustness of higher-rank symmetry breaking against disorder, which claims that the quantum
claim that the quantum critical dimension in the presence of disorder
remains the same as the classical critical dimension. 
Remarkably, the proof of the generalized Mermin-Wagner theorem to the dipole symmetry
was    provided in \citet{PhysRevB.106.245125} for a system with  finite-range interactions
and clustering. They found that the dipole symmetry cannot be spontaneously broken
if $d=1,2$ and $T>0$. One direct consequence is that a system of
fermions with a microscopic dipole symmetry cannot flow to a Fermi
liquid or any phase with a pronounced Fermi surface.

\section{Higher-rank symmetry defects and Kosterlitz-Thouless phase transitions\label{section_kt}}

\begin{figure}
\centering \includegraphics[scale=0.6]{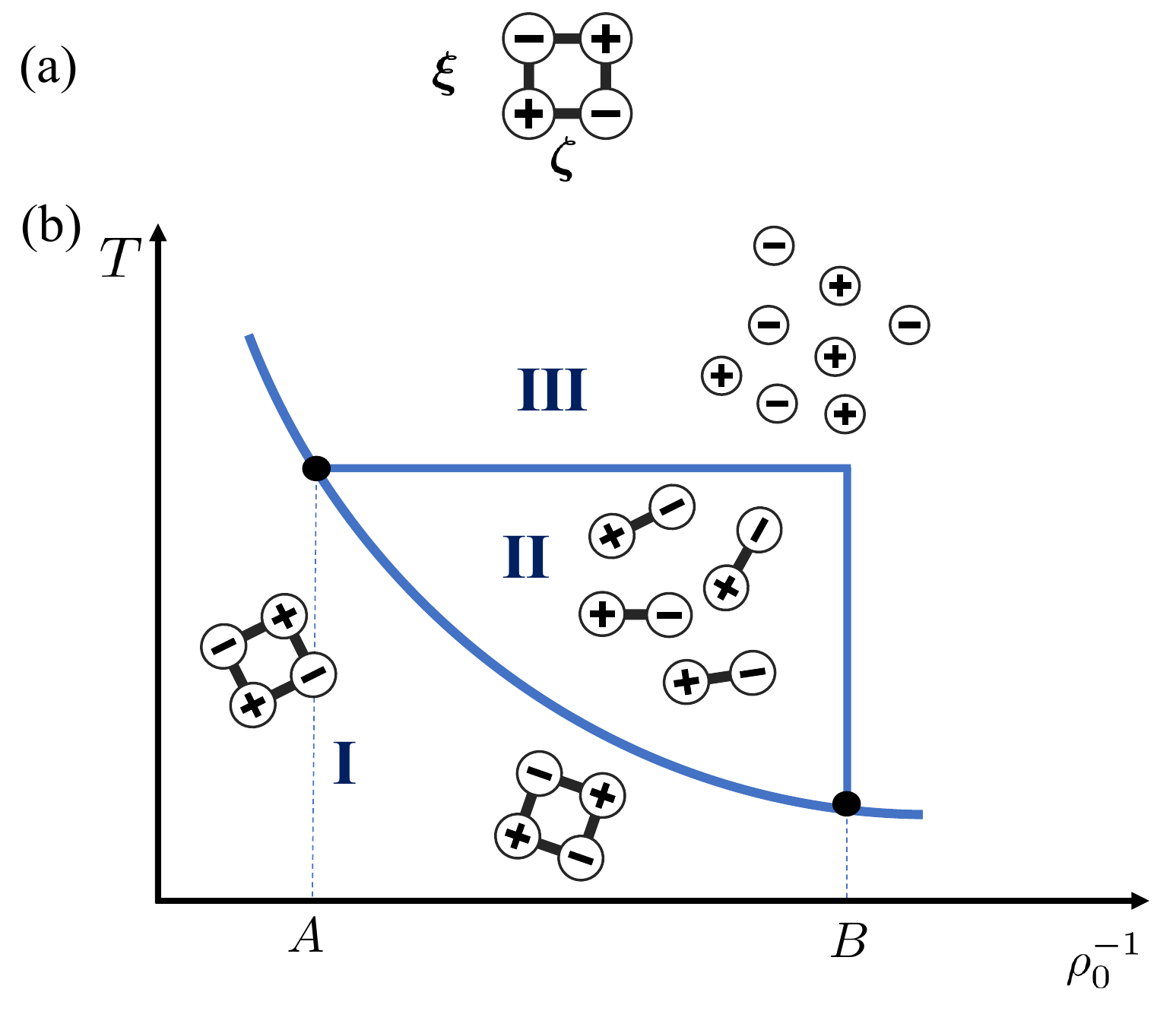} \caption{
Global phase diagram for the minimal model that preserves the angular moment. (a) A quadrupole bound state of a
pair of a unit defect and a unit anti-defect. (b) Finite-temperature
phase diagram from the renormalization group analysis and DebyeHuckel approximation. $T$ and $\rho^{-1}_0$
are respectively temperature and superfluid stiffness inverse. In Phase-I, defects $\Theta$ are confined in the
form of quadrupole bound states. In Phase-II, defects are confined
in the form of dipole bound states. In Phase-III, defects are fully
released from bound states and are wildly mobile. The locations of
A and B depend on core energy values of both defects and dipole
bound states. Two topological transitions occur
successively from Phase-I, Phase-II, to Phase-III. A direct transition
from Phase-I to Phase-III occurs if $\rho^{-1}_0$
is outside the domain (A, B).
From~\protect\citet{2022arXiv220108597Y}.}
\label{Fig:phasediagram} 
\end{figure}


\subsection{Construction of higher-rank symmetry defects}\label{section_kt_construction}

The complicated feature of the higher-rank symmetry will enable a
more complicated defect structure. The
thermal vortices are fundamental to a superfluid phase in congruent
with the gapless modes. The existence of higher-rank symmetry admits
a complicated structure as defects should inevitably inherit the structure
from the higher-rank group. For the higher-rank defects, one way is to apply the duality method by studying the structure of monoples in a tensor gauge field. 
Here we introduce a new systematical way to construct the defects \citep{2020arXiv201003261C,2022arXiv220108597Y}.
Mathematically, one may use a multi-valued
phase function to characterize the defects. Given a phase field $\theta$,
it can be always decomposed as the smooth part $\theta^{s}(\mathbf{x})$
and the multi-valued part $\theta^{v}(\mathbf{x})$. In general, in
two spatial dimensions, the multivalued component $\theta_{a}^{v}(\mathbf{x})$
can be formulated as $\theta_{a}^{v}(\mathbf{x})=f_{a}(\mathbf{x})\varphi(\mathbf{x})$,
where $\varphi(\mathbf{x})$ defined mod $2\pi$ is the angle of site
$\mathbf{x}$, which an equivalent relation $\theta_{a}^{v}(\mathbf{x})\sim\theta_{a}^{v}(\mathbf{x})+2\pi f_{a}(\mathbf{x})$.
Subtly, $f_{a}(\mathbf{x})$ should be validated under a lattice regularization
to protect single valuedness of field $\phi_{a}$, where spatial coordinates
$\mathbf{x}$ are regarded as $\mathbf{x}=(x^{1},x^{2})=\mathbf{n}a$
with $\mathbf{n}=(n_{1},n_{2})$ being a pair of integers and $a$
being the lattice constant. The equivalence relation resembles gauge
freedom. That is, whether we start with $\theta_{a}^{v}(\mathbf{x})$
or $\theta_{a}^{v}(\mathbf{x})+2\pi f_{a}(\mathbf{x})$ should cause
no physical effects.This leads to the Statement~\ref{state1}. 
\begin{statement}
The physical Hamiltonian density should be single valued even in the
presence of multivalued vortex configurations. 
\label{state1}
\end{statement}
The Statement~\ref{state1} clarifies the Hamiltonian density $\mathcal{H}[\theta_{a}(\mathbf{x})]$
is invariant when $\theta_{a}(\mathbf{x})$ is shifted by $2\pi f_{a}(\mathbf{x})$,
$\mathcal{H}[\theta_{a}(\mathbf{x})+2\pi f_{a}(\mathbf{x})]=\mathcal{H}[\theta_{a}(\mathbf{x})]$,
which determines the most singular part of a vortex. Take a conventional
superfluid as an example with Hamiltonian density $\mathcal{H}=\frac{1}{2}[(\partial_{1}\theta)^{2}+(\partial_{2}\theta)^{2}]$.
With the assumption $\theta^{v}(\mathbf{x})=f(\mathbf{x})\varphi(\mathbf{x})$,
the constraint imposed by Statement 1 on shifting $2\pi f(\mathbf{x})$,
gives the equations $\partial_{1}f(\mathbf{x})=0,\partial_{2}f(\mathbf{x})=0$
towards which we have the solution $f(\mathbf{x})=\ell$ with $\ell\in\mathbb{Z}$.
Thus we recover vortex configurations in a conventional superfluid.

The smooth component is controlled by the second statement once we
obtain the multivalued component. For convenience, one may conventional
$U(1)$ charges as rank-0 while the others are higher-rank charges. 
\begin{statement}
The action of a higher-rank symmetry group on some bound states of
operators charged in the higher-rank symmetry group is equivalent
to an action of a global $U(1)$ symmetry with appropriate rank-$0$
charges. \label{state2}
\end{statement}
The Statement~\ref{state2} allows us to construct a set of bound states such
that the higher-rank group only induces a global phase shift. Explicitly,
given defects carrying higher-rank charges, the Statement~\ref{state2} claims
that some bound state of these vortices is proportional to $\varphi(\mathbf{x})$
bearing the behavior of a conventional vortex, that is, the smooth
component vanishes. Thus the essence is to find the bound states that
behave like a conventional defect.

One can straightforwardly apply the two construction statements to
the two minimal models \citep{2020PhRvR2b3267Y,2020arXiv201003261C} (also see Table~\ref{Tab}). For the model of the dipole conservation,
there are two kinds of defects: the conventional one with $\theta=\ell\varphi(\mathbf{x})$,
while the unconventional ones with 
\begin{equation}
\theta=p_{1}x^{1}\varphi(\mathbf{x})+p_{1}x^{2}\log|\mathbf{x}|
\end{equation}
or 
\begin{equation}
\theta=p_{2}x^{2}\varphi(\mathbf{x})-p_{2}x^{1}\log|\mathbf{x}|.
\end{equation}
Here $\ell\in\mathbb{Z}$ represents the winding number and $p_{1,2}$
has the dimension $[x]^{-1}$, which may be dubbed a dipole charge.
For the minimal model of angular moment conservation, the conventional
defects take a dipole charge $p$ with a configuration 
 \begin{align}
\theta_{1}\left(\mathbf{x}\right)& =-px^{1}\log|\mathbf{x}|+px^{2}\varphi\left(\mathbf{x}\right)\label{eq:defect1}  \\
\theta_{2}\left(\mathbf{x}\right) & =-px^{2}\log|\mathbf{x}|-px^{1}\varphi\left(\mathbf{x}\right) \label{eq:defect2}
\end{align}
while conventional defects can be regarded as bound states of these
unconventional defects. One may notice that the charge $p$ should
be regularized as $p=\ell a^{-1}$ $\left(\ell\in\mathbb{Z}\right)$.
When we circle around the vortex core, the vortex configuration get
an extra phase $\delta\theta_{1}=2\pi px^{2}=2\pi\ell n_{2}$ and
$\delta\theta_{2}=-2\pi px=-2\pi\ell n_{1}$ with $\mathbf{x}=(n_{1},n_{2})a$,
which keeps in consistence with compactness of $\theta_{a}$.

 \subsection{Hierarchy of Kosterlitz-Thouless phase transitions}\label{section_kt_rg}

The $2\mathsf{SF}^1$ has an true ODLRO with non-vanishing
order parameters \citep{2020arXiv201003261C}, with the low-energy effective Hamiltonian in Eq.~\eqref{eq:eff2sf1} or
$\mathcal{H}[\theta_1,\theta_2]=\frac{1}{2}\rho_{0}[(\partial_{1}\theta_{1})^{2}+(\partial_{2}\theta_{2})^{2}+(\partial_{1}\theta_{2}+\partial_{2}\theta_{1})^{2}]$,
where $\theta_{1,2}$ are Goldstone modes.
 Here for convenience, we use $\Theta $ to represent the defect configurations  in 
Eqs.~(\ref{eq:defect1}) and (\ref{eq:defect2}):
 \begin{equation}
\!\!\!\!\!\Theta\equiv(\theta_{1,}\theta_{2})\!=\!\ell\left(\frac{x^2}{a}\varphi(\mathbf{r})-\frac{x^1}{a}\ln\frac{r}{a},-\frac{x^1}{a}\varphi(\mathbf{r})-\frac{x^2}{a}\ln\frac{r}{a}\right).\!\!\!\label{rank-1}
\end{equation}
and a conventional defect can be taken as a bound state of $\Theta$.
  The finite temperature phase diagram is shown in Fig.~\ref{section_kt}.
A preliminary understanding can be reached by inspecting the total energy $\mathscr{E}[\{\Theta_i\}]$ of $N$ defects $\{\Theta_i\}$ carrying charges $\ell_{i}$
with the core coordinates $\mathbf{r}_{i}$ ($i=1,\ldots,N$),
\begin{align}
\mathscr{E}[\{\Theta_i\}]=\frac{\rho_{0}}{2}\sum_{i,j=1}^{N}\ell_{i}\ell_{j}\int d^{2}\mathbf{q}\mathscr{U}_\mathbf{q}e^{i\mathbf{q}\cdot(\mathbf{r}_{i}-\mathbf{r}_{j})}
\end{align}
where $\mathscr{U}_\mathbf{q}=\frac{2}{q^{4}}$ and the momentum $q=|\mathbf{q}|$. 
  In this expression, a series of severely infrared divergent terms can be identified. By observing
\begin{equation}
\!\!\!\!\int \!d^{2}\mathbf{q}\mathscr{U}_\mathbf{q}e^{i\mathbf{q}\cdot(\mathbf{r}_{i}-\mathbf{r}_{j})}\!=\!2\pi\bigg(\frac{L}{a}\bigg)^{2}\!- \pi\frac{\vert\mathbf{r}_{i}-\mathbf{r}_{j}\vert^{2}}{a^{2}}\ln\!\frac{L}{a}+\! f,\label{eq:U11}
\end{equation}
we can single out two types of divergence, i.e., $L^{2}$ and $\ln L$,
while the letter $f$ incorporates all finite terms. In the low temperature
region, we have to impose the constraints, the charge neutral condition and dipole moment neutral condition, on the defect configuration
for energetic consideration to cancel the $L^{2}$ and $\ln L$ divergence, respectively.
 Therefore, there are basic three distinct objects, namely, a single defect, a dipole bound state (BCS), and a quadruple BS. A single defect and dipole BS possess a divergent self-energy while a quaduple BCS that formed by four defects with vanishing charges and vanishing dipole moments as shown in Fig.~\ref{Fig:phasediagram} has a finite self-energy.

Bearing the above analysis in mind, one can conclude that, at the thermodynamic limit, HRS defects are energetically
confined into quadruple BSs at low temperatures (Phase-I) in Fig.~\ref{Fig:phasediagram}. Phase-I shows the algebraic long-range order since the correlation function $\langle \hat\Phi(\mathbf r_1)\hat\Phi^\dagger(\mathbf r_2)\rangle$ behaves as a power-law function at long distances since the low-energy physics is dominated by gapless phonon mode excitations with a renormalized stiffness by defects.
In phase II in Fig.~\ref{Fig:phasediagram}, defects are confined in dipoles.
dipole BSs will be completely released from quadrupole BSs when the temperature 
goes beyond the critical value $T_{c1}$. 
In Phase-III, defects $\Theta$ are released and deconfined.
When the temperature is close to the second topological phase transition  $T_{c2}$
from Phase-II to Phase-III,
dipole BSs form a plasma due to  sufficiently high density and equivalently we 
have a condensate of dipole BSs. 
Therefore, the interaction between two defects $\Theta$  gets strongly renormalized.
This dynamical renormalization can be quantitatively treated via the standard  
Debye-H{\"u}ckel approximation \citep{chaikin2000principles} by approximating the dipole BSs as a continuous independent field. 
There is another possible transition from Phase I directly to Phase III,
due to negative screened core energy of a single defect $\Theta$ or $T_{c2}<T_{c1}$. 
In the former case, instead, the screening
effect favors a finite density of defects $\Theta$, with a vertical line as a boundary between Phase II and Phase III. In the latter case, Phase II is metastable and the two transitions merge together.
In both cases, we end up with  a direct transition  from Phase-I to Phase-III, where  quadruple BSs are directly dissolved into free isolated defects. 
At the high temperature limit in Phase-III where defects $\Theta$
reach a sufficiently high density, a plasma phase forms and 
the interaction between $\Theta$ will get screened 
from $\frac{2}{q^{4}}$ to $\frac{2}{q^{4}+\lambda_{s}^{-4}}$
with $\lambda_{s}$ temperature-dependent Debye-H\"uckel screening length
of defects $\Theta$.  

\section{Hydrodynamics governed by Navier-Stokes-like equations}\label{section_hydro}
Hydrodynamics appears as a powerful tool in describing a 
many-body dynamics and thermalization, which applies to both 
quantum and classical system. As expected, the fractonic superfluid 
should exhibit exotic hydrodynamical behaviors due to a higher-rank symmetry.
In this section, we introduce the highly
unconventional hydrodynamics of a fractonic superfluid that is exemplified
by the minimal system that preserves the angular moment in Sec.~\ref{section_minimal_model_angular}.

We can derive the hydrodynamical equation from the EL equation from the  by splitting
the field $\phi_{a}$ into the amplitude and the phase \citep{2022ChPhL..39e7101Y}. Accordingly
we can define the velocity field $v_{ab}$ Neother currents according to Eqs.~\eqref{u(1)current} and \eqref{dipolecurrent}. Then the Navier-Stokes-like
equations for the diagonal components $v_{aa}$ ($a=1,\cdots,d$)
can be written in a more compact form $\partial_{t}v_{aa}=\partial_{a}(-T_{a}+p_{a})$,
which takes the similar form as a conventional superfluid. For the
off-diagonal components $v_{ai}$ ($i\neq a$), we have 
\begin{equation}
\partial_{t}v_{ai}=K_{ai}\rho_{i}[\partial_{i}(-T_{a}+p_{a})+\partial_{a}(-T_{i}+p_{i})]-\frac{v_{ai}}{\rho_{i}}\sum_{b}\partial_{b}(\rho_{i}v_{ib}),
\end{equation}
where
\begin{align}
T_{a}=\frac{v_{aa}^{2}}{2}+\sum_{b(\neq a)}\frac{v_{ab}^{2}}{2K_{ab}\rho_{b}}
\end{align}
is the kinetic density for a-component and 
\begin{align}p_{a}=\frac{1}{2\sqrt{\rho}}\sum_{j}\partial_{j}^{2}\sqrt{\rho}
\end{align}
is the quantum pressure term from lineon effects. The kinetic density
of immovable direction $x^{b}$ contains the $b$-component density
field $\rho_{b}(\mathbf{x})$, and may characterize the difference between
fractonic hydrodynamics and conventional hydrodynamics. For the off-diagonal
components, the density $\rho_{i}\left(\mathbf{x}\right)$ outside
the partial derivative makes currents in the immovable direction beyond
the conventional Navier-Stokes equation. However, for the ground state
of $d\mathsf{SF}^{1}$, i.e. $\rho_{0}=\mu/g$ and $\theta_{a}^{\mathrm{cl}}$,
the pressure term $p_{a}\equiv0$ and the off-diagonal components
can be further simplified: $\partial_{t}v_{ai}=-K_{ai}\rho_{0}(\partial_{i}T_{a}+\partial_{a}T_{i})-v_{ai}\sum_{b}\partial_{b}v_{ib}$. While
the the first two terms are more familiar, the physical meaning of
the third term is still an open question.

The general discussion on the hydrodynamics of systems with higher moment conservation is a focus of study recently, see, e.g.,   \citet{2020PhRvR2c3124G,2021PhRvR...3d3186G,PhysRevB.100.214301,2021PhRvD.104j5001A,2023JHEP...03..188H,2023arXiv230102680G,2023arXiv230309573S}.  
In a conventional superfluids the Goldstone boson of the broken symmetry
becomes a hydrodynamic mode. However, 
exotic features will arise due to the implementation of higher-rank symmetries.
For example \citep{2023arXiv230309573S}, given a system of both global U(1) charge and dipole moment,
there is quadratic subdiffusion in absence of     SSB. While both charge and 
dipole symmetries are broken, there is a quadratically mode. 
In the phase only with dipole symmetry broken and global U(1)
charge preserve, there are both diffusive
transverse modes and longitudinal modes and the later can be 
either purely diffusive or quadratically propagating depending on parameters.

\section{Realization in lattice models}\label{section_latt}

\begin{figure}
\includegraphics{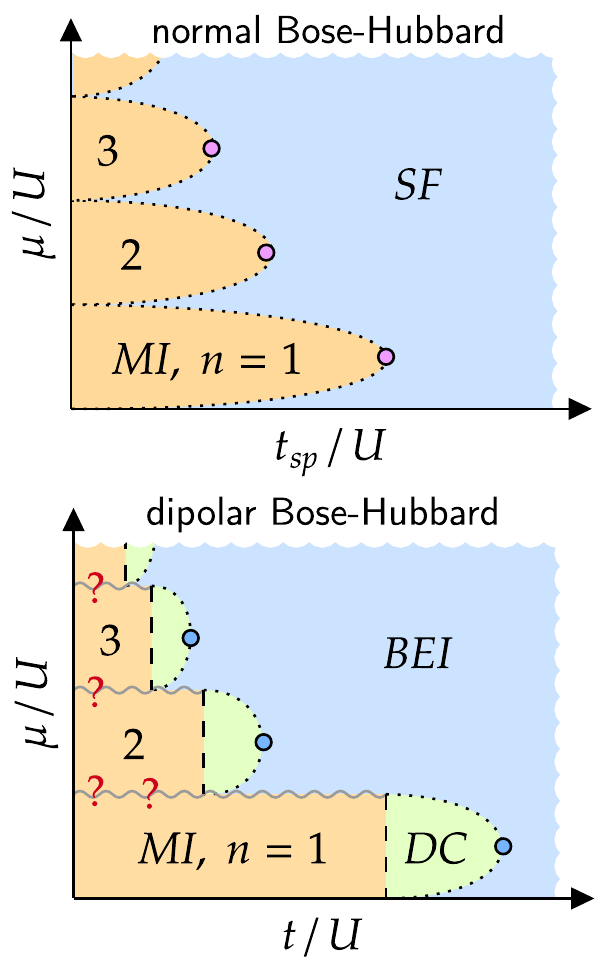} 
\caption{\label{fig:phase_diagram} Mean-field quantum phase diagram of the
normal Bose-Hubbard model (top) and the dipolar Bose-Hubbard model
(bottom). The orange regions are Mott insulators, with the integer
labels denoting the average density of bosons on each site. The green
regions are dipole condensates, where dipoles (but not single bosons)
have condensed; the average density of these phases is the same as
that of their parent Mott insulators. The blue region in the top plot
is a conventional superfluid, while the blue region in the bottom
plot is a `Bose-Einstein Insulator' or equivalently fractonic superfluid 
phase denoted by $d\mathsf{SF}^{0}$. From~\protect\protect\protect\citet{2022PhRvB.106f4511L}. }
\end{figure}

There are various types of lattice model realization of many-fracton systems (see, e.g., \citet{2020PhRvL.125x5303F,2022PhRvR...4b3151G,2022PhRvB.106s5139G}).
One of the lattice realization approaches is the dipolar Bose-Hubbard
model \citep{PhysRevB.40.546,2022PhRvR...4b3151G,2022PhRvB.106f4511L,2022arXiv221002470L}, which describes
interacting bosons hopping on a d-dimensional cubic lattice in a manner
that conserves both total boson number and total boson dipole moment.
The Hamiltonian $H_{\mathrm{DBHM}}=H_{\mathrm{hop}}+H_{\mathrm{onsite}}$
is the modified Hubbard model on a square lattice 
with hopping terms that respects the dipole symmetry, 
\begin{align}
H_{\mathrm{hop}} & =-\sum_{i,a}(tb_{i-a}^{\dagger}b_{i}^{2}b_{i+a}^{\dagger}+t^{\prime}\sum_{b\neq a}b_{i}^{\dagger}b_{i+a}b_{i+a+b}^{\dagger}b_{i+b}+h.c.)\,,\\
H_{\mathrm{onsite}} & =\sum_{i}\left[-\mu n_{i}+\frac{U}{2}n_{i}(n_{i}-1)\right]\,,
\end{align}
where $t,t^{\prime},\mu,U$ are all positive coefficients and $n_{i}=b_{i}^{\dagger}b_{i}$
is the boson number operator on site $i$. 
with the hopping terms capturing the simplest boson $b_{i}$ hopping
processes compatible with dipole conservation. Thus, there are two
types of fundamental particles: the mobile dipole $d_{j}^{a}\equiv b_{j+a}^{\dagger}b_{j}$
and the localized single boson $b_{j}$.What lies at the heart is
the competition between $H_{\mathrm{hop}}$ and $H_{\mathrm{onsite}}$.
The mean-field field solution predicts a fruitful phase diagram as shown in Fig.~\ref{fig:phase_diagram}. In
particular, the first phase is where the neutral objects carrying
nonzero dipole moment condense, with the gap to charged excitations
remaining nonzero across the transition. Such dipole condensation
is natural since the symmetry constrains the isolated doped particles
to be localized while the dipolar bound states can move freely. A
second phase $2\mathsf{SF}^{0}$ arises due to condensation of the single bosons that lack mobility,
which turns out to be an insulator phase due to the absence of the Meissner
effect and the constraint from the dipole conservation. The author
 calls it the \textit{Bose-Einstein insulator} (BEI) owing to the absence of Meissner effect,
 which can be attributed to the constraint from the dipole conservation
 on the charge current, 
 while the effective theory for the Goldstone
 mode is an anisotropic quantum Lifshitz model that is the same as
 the one in fractonic superfluid.

As we know, the presence of global conserved quantities in interacting
systems generically leads to diffusive transport at late time. In
contrast, the system conserving the dipole moment of the associated
global charge, or even higher moment generalizations will bear subdiffusive
decay. For dipole-conservation, the modeling simulations provide
a postdiction of the subdiffusive scaling experimentally observed
in~\citet{PhysRevB.101.174204}.

\section{Outlook}\label{section_discussion}

The investigation of higher-rank symmetry in quantum many-fracton systems has garnered significant interest and attention from researchers in recent years. 
The intertwine between spontaneous symmetry breaking and many-fracton systems, where mobility constraints limit the motion of constituent particles, has opened up new possibilities for engineering novel quantum phases of matter using higher-rank symmetries.
For instance, researchers may explore the possibility of creating pair density wave or FFLO phases inspired by the momentum-carrying ground state in fractonic superfluids. 
Moreover, it remains a significant challenge to find a mathematical tools to classify 
the higher-rank defect.
For instance, the hierarchical structure that emerges from higher-rank symmetry could potentially enrich homotopy groups that successfully characterize defects in a conventional superfluid phase. 
Additionally, it may be speculated that these higher-rank symmetry enriched phase may lead to the emergence of new universality classes of quantum phase transition.
Analogous to the symmetry protected topological phases, one may envision 
the higher-rank symmetry protected phase possibly with exhibit exotic properties, which may exert new insights into the behaviors of quantum many-body systems.
In conclusion, the study of higher-rank symmetry in quantum many-fracton systems is an ever-evolving and fascinating domain, offering a plethora of possibilities for creating new and unconventional phases of matter. 
 
%


\end{document}